\documentclass[useAMS,usenatbib]{mn2e}

\usepackage{graphicx}
\usepackage{amsmath}

\title[Orientation and quasar black hole mass estimation]{Orientation and quasar black hole mass estimation}
\author[Brotherton, Singh, \& Runnoe]{Michael S. Brotherton$^{1}$\thanks{E-mail: mbrother@uwyo.edu}, Vikram Singh$^{1}$, Jessie Runnoe$^{2}$ \\
$^{1}$Department of Physics and Astronomy, University of Wyoming, Laramie, WY 82071, USA\\
$^{2}$Department of Astronomy and Astrophysics, The Pennsylvania State University\\ 525 Davey Lab, University Park, PA 16803, USA\\
}
\begin{document}		

\date{}

\pagerange{\pageref{firstpage}--\pageref{lastpage}} \pubyear{2015}

\maketitle

\label{firstpage}

\begin{abstract}

We have constructed a sample of 386 radio-loud quasars with $z < 0.75$ from the 
Sloan Digital Sky Survey in order to investigate orientation effects on 
black hole mass estimates.  Orientation is estimated using radio core dominance
measurements based on FIRST survey maps. Black hole masses are estimated from
virial-based scaling relationships using H$\beta$, and compared to 
the stellar velocity dispersion ($\sigma_*$), predicted using 
the Full Width at Half Maximum (FWHM) of [O III] $\lambda$5007, 
which tracks mass via the M-$\sigma_*$ relation. 
We find that the FWHM of H$\beta$ correlates significantly
with radio core dominance and biases black hole mass determinations that
use it, but that this is not the case for $\sigma_*$ based on [O III] $\lambda$5007.  
The ratio of black hole masses predicted using orientation-biased 
and unbiased estimates, which can be determined for radio-quiet as well as radio-loud quasars, 
is significantly correlated with radio core dominance.  Although there is significant scatter, 
this mass ratio calculated in this way may in fact serve as an orientation estimator.  
We additionally note the existence of a small population
radio core-dominated quasars with extremely broad H$\beta$ emission lines 
that we hypothesise may represent recent black hole mergers.

\end{abstract}


\section{Introduction}

Quasars are not spherically symmetric systems.  This is most apparent
in the large-scale radio jets present in a minority of sources,
but several inferred quasar structures, notably accretion disks and dusty torii, 
are also axisymmetric.  In radio-loud quasars, radio core dominance and 
spectral indices can be used to estimate system orientation (Orr \& Browne 1982;
 Wills \& Brotherton 1995; Van Gorkom et al. 2015).
There is as yet no reliable orientation indicator for radio-quiet quasars, which represent
about 90\% of all quasars (but see Boroson 2011 and Sulentic et al. 2012).

This axisymmetry is important to understand and account for when estimating
some fundamental quasar parameters, such as black hole mass.
Many methods of determining quasar black hole mass (e.g., Shen 2013, Peterson 2011) 
depend on the assumption of virial motion and use the Doppler-broadended velocity 
width of an emission line -- usually H$\beta$ at low redshift.  The broad
H$\beta$ emission-line width, often characterized by the Full Width at 
Half Maximum or FWHM, however, is correlated with orientation
indicators in radio-loud quasars (e.g., Wills \& Browne 1986). This 
is suggestive that at least part of the broad-line region (BLR) is
flattened into a disk, with edge-on systems showing the broadest lines,
a conclusion also supported by other lines of evidence (Gaskell 2009a).  
The velocity width of lines like H$\beta$ are apparently related to both black hole
mass and system orientation.

Runnoe et al. (2013) compared black hole masses computed using single-epoch
(SE) scaling relationships (e.g., Vestergaard \& Peterson 2006) for both
H$\beta$ and C IV $\lambda$1549, the latter lacking strong FWHM changes
with orientation (but see Runnoe et al. 2014 for a discussion of some subtle profile changes).  
They found that the difference in the two black hole mass estimates significantly correlated with
radio-based orientation parameters and determined correction terms for the 
H$\beta$-based formula.  The same orientation effects affecting H$\beta$ 
also appear to be present in radio-quiet quasars (Shen \& Ho 2014). 
It would be valuable to be able to provide an estimation of orientation
for all quasars, not just radio-loud objects, to better determine
how quasar properties change with viewing angle.

Another method exists for estimating black hole mass in quasars,
via the correlation with stellar velocity dispersion $\sigma_*$
(e.g., Gebhardt et al. 2000; Ferrarese \& Merritt 2000).  The so-called 
M-$\sigma_*$ relation, present in both active and quiescent galaxies,
may represent an important clue to galaxy evolution.
In the majority of quasars, the continuum dilutes the stellar absorption
features to the point of non-detection, making $\sigma_*$ difficult to
determine directly.  Shields et al. (2003) pioneered using the 
FWHM of [O III] $\lambda$5007 (divided by 2.35 under the assumption
of a Gaussian profile) as a proxy for $\sigma_*$ in luminous quasars.
Others have since verified that FWHM [O III] $\lambda$5007 does 
correlate, albeit with significant scatter, with $\sigma_*$ (Boroson 2003; 
Gaskell 2009b), although the [O III] profile is not generally Gaussian 
and the relationship is not so simple.   Brotherton (1996) found
no relationship between FWHM [O III] $\lambda$5007 and radio-based
orientation indicators, although $\sigma_*$ 
measurements may depend on the orientation to the line of sight
of host galaxies disks (e.g., Woo et al. 2015).

Other investigations by Salviander et al. (2008) and Salviander \& Shields (2013) 
used an H$\beta$-based scaling relationships to determine the quasar black hole
mass, and FWHM [O III]/2.35 as a proxy for $\sigma_*$, in order to study the 
M-$\sigma_*$ relation using large samples of spectra from the
Sloan Digital Sky Survey (SDSS).  Notably, while they found M/$\sigma_*$ significantly 
correlated with redshift, they accounted for most of the effect 
as the result of a variety of selection biases and argued that
M$_{\rm BH}$-$\sigma_*$ did not strongly evolve.  They did not identify 
differences between radio-loud and radio-quiet quasars.  
We propose that much of their scatter in M$_{rm BH}$-$\sigma_*$ arises 
from the orientation bias in FWHM H$\beta$, and 
that it may be possible to use scatter in the M$_{rm BH}$-$\sigma_*$ relation
computed in this way as an orientation indicator.

In the rest of this paper, we first recalibrate
FWHM [O III] $\lambda$5007 as a predictor of $\sigma_*$ without assuming 
a Gaussian profile (\S 2.1).  Then we construct a radio-loud
quasar sample using the SDSS for which we can measure M$_{\rm BH}$-$\sigma_*$
and compare it to radio-core dominance, an orientation indicator (\S 2.2).
We calculate the correlation matrix for quantities of interest (\S 3),
finding strong orientation effects as expected.  We discuss
prospects for, and challenges with, using our results to estimate orientation
for quasars in general (\S 4).  We also discuss the nature of a surprising 
new class of outliers having extremely broad emission lines but large radio core
dominance that may be merger products (\S 4).
Finally, we summarize our conclusions (\S 5).

\section{Samples, Data, and Methods}


\subsection{Calibration of FWHM$_{\rm [O III]}$ as a $\sigma_*$ predictor}

Instead of assuming a profile shape for the narrow [O III] $\lambda$5007 emission
line or relationship with the stellar velocity dispersion, we prefer
to empirically calibrate FWHM$_{\rm [O III]}$ as a predictor of  $\sigma_*$.
This has been previously done (e.g., Boroson 2003, Gaskell 2009b), but 
we can now do this with much larger samples with the same quality of spectra that 
we will work with throughout this investigation.

We started with the sample of Shen et al. (2008), who measured $\sigma_*$ from
stellar absorption lines in more than 900 broad-lined AGNs with spectra
from the Sloan Digital Sky Survey (SDSS).  Their measurements included 
corrections for instrumental resolution and aperture effects.
They did not make any measurements of the narrow [O III] profiles, however.

We downloaded SDSS spectra for these objects and made our own measurements.
Along with a local continuum and an appropriately broadened Fe II template
(Boroson \& Green 1992), 
we fit [O III] $\lambda$5007 and [O III] $\lambda$4959 simultaneously with a fixed 
3:1 intensity ratio, using the same profile comprised of the sum of two 
skewed Gaussians with different peak wavelengths allowed to vary.  This mathematical function
satisfactorily reproduced the full range of observed [O III] profiles in the sample.
From these analytical fits, we numerically determined the FWHM.
We obtained the instrumental resolution at the wavelength of [O III] $\lambda$5007
from the SDSS headers and subtracted this in quadrature from the observed FWHM to 
determine an intrinsic FWHM. This is an important step because the 
smallest FWHM values are not much larger than the intrinsic spectral
resolution.

We determined errors using an empirical formula based 
on Monte Carlo simulations.  We created a grid of artificial [O III]
profiles with a range in signal-to-noise ratios and equivalent widths
spanning those seen in our data set. For each grid point, we created 50
artificial spectra with random Gaussian noise added at different levels and
measured profile properties. The distribution of measurements provides
the uncertainties for the grid.  We then employed multiple regression using
signal-to-noise ratio and equivalent width as predictors of the 
uncertainty and used the resulting fit to determine individual error bars.
The mean uncertainty on our final FWHM [O III] measurements was 11$\pm$3\%.


We excluded a number of objects for several reasons.     
Upon visual inspection, 
some spectra had poor signal-to-noise ratios or other spectral defects in the
region of [O III] compromising our fits.  Some spectra show double-peaked [O III] profiles,
which are of interest as binary candidates (e.g., Fu et al. 2011), but which are unlikely
to be representative of typical line profiles and might skew our work.  A few spectra would not
properly download from the SDSS servers.  Our final sample consists of 773 objects.
Table 1 provides our new measurements of FWHM [O III] along with $\sigma_*$ with
uncertainties and aperture corrections from Shen et al. (2008).

\begin{table*}
 \centering
 \begin{minipage}{140mm}
  \caption{Measurements for the revised Shen et al. (2008) sample.  The full table is available online.}
  \begin{tabular}{@{}ccccccc@{}}
  \hline
SDSS Name & z & $\sigma_*$ & Corr $\sigma_*$ & FWHM [O III] & Inst. res & Intrinsic FWHM [O III] \\
(J2000)  & & (km s$^{-1}$) & (km s$^{-1}$) & (km s$^{-1}$) & (km s$^{-1}$) & (km s$^{-1}$) \\
\hline
000611.54+145357.2 & 0.1186 & 181.1$\pm$13 & 189 & 344.8 & 136.2 & 316.8$\pm$36 \\
000729.98$-$005428.0 & 0.1454 & 121.2$\pm$18 & 128 & 269.3 & 132.3 & 234.6$\pm$27 \\
000805.62+145023.3 & 0.0455 & 173.1$\pm$8 & 173 & 407.6 & 146.1 & 380.6$\pm$38 \\
000813.22$-$005753.3 & 0.1393 & 221.8$\pm$18 & 233 & 475.8 & 133.1 & 456.8$\pm$36 \\
001255.02+010905.2 & 0.1547 & 140.5$\pm$24 & 148 & 311.8 & 143.0 & 277.1$\pm$33 \\
001340.21+152312.0 & 0.1196 & 92.4$\pm$14 & 96 & 265.2 & 128.0 & 232.2$\pm$30 \\
001903.17+000659.1 & 0.0727 & 118.9$\pm$8 & 121 & 252.6 & 139.0 & 210.8$\pm$27 \\
001954.35+155740.3 & 0.0828 & 98.8$\pm$9 & 101 & 259.2 & 127.9 & 225.5$\pm$28 \\
002305.03$-$010743.4 & 0.1661 & 170.5$\pm$18 & 180 & 364.0 & 138.1 & 336.7$\pm$38 \\
002839.65+145138.8 & 0.1963 & 69.4$\pm$17 & 74 & 238.8 & 144.3 & 190.4$\pm$30 \\
\hline
\end{tabular}
\end{minipage}
\end{table*}

Figure 1 plots log FWHM [O III] against the log of the aperture-corrected stellar velocity dispersion $\sigma_*$.
Objects with radio flux densities greater than 10 mJy at 1.4 GHz 
with the NRAO VLA Sky Survey (NVSS, Condon et al. 1998) and within 30$\arcsec$ of the SDSS
position (as recommended by Kimball \& Ivezi{\'c} 2008) are distinguished by filled circles.
With only 17 objects, the filled circles do not differ at a statistically significant level from the open circles, 
and we conclude that there is no strong dependence on radio-loudness.
We also note that [O III] profiles are not Gaussians, and that
FWHM [O III]/2.35 generally underpredicts $\sigma_*$, as shown
on the plot, and should not be used.

The quantities log FWHM [O III] and log $\sigma_*$ are strongly correlated, 
with a Spearman correlation coefficient of 0.53, which is extremely significant with 773 points. 
Care must be taken when choosing a method of fitting a line in the presence of scatter, 
keeping in mind the intended application of the fitted relationship.
The appropriate line fit to predict one parameter from another
in an unbiased manner is the ordinary least squares (OLS) Y on X line (e.g., Isobe et al. 1990). 
Unbiased here means that the residuals between the y parameter and the fitted line 
do not correlate with the x parameter.
The OLS Y on X line shown on to predict log $\sigma_*$ based on log FWHM [O III], all
in km s$^{-1}$ is:

\begin{equation}
\begin{aligned}
{\rm log\ } \sigma_{* [OIII]} = & 1.0366(\pm0.0712) + \\
& 0.4701(\pm0.0284) {\rm log\ FWHM [O III]}
\end{aligned}
\end{equation}

Extra digits are provided in the fitting parameters for plotting purposes.
This produces a distribution of y-axis residuals centered on zero as desired
for prediction, with a standard deviation of 0.12 dex, or about 30\%.

The OLS Y on X line, especially when significant scatter is present, 
can differ from more symmetric approaches to fitting.
A line fit that treats the x and y axes symmetrically is 
more appropriate when trying to determine the 
underlying relationship between two variables.  Given that this may
be of interest, we also provide the OLS Y on X and OLS X on Y 
bisector fit:

\begin{equation}
\begin{split}
{\rm log\ } \sigma_* = &  -0.12252(\pm0.0323) \\
& + 0.9332(\pm0.0126) {\rm log\ FWHM [O III]}
\end{split}
\end{equation}

We generated the $1 \sigma$ uncertainties for the above equation with a Monte Carlo approach,
using the observed parameter distributions and 1000 iterations.

\begin{figure}
\begin{center}
\includegraphics[width=8.9 truecm]{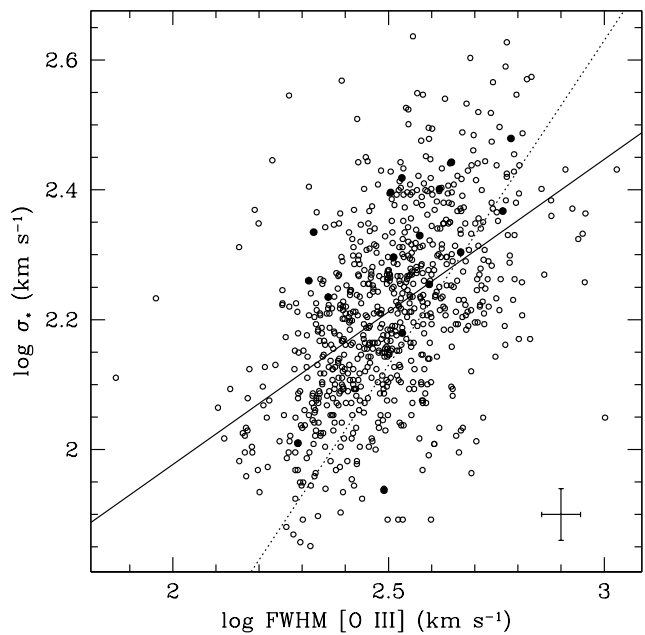}
\end{center}
\caption{
We plot log FWHM [O III] against log $\sigma_{* [OIII]}$ for our final sample
of 773 objects, with the filled circles indicating the 17 sources
detected in the radio with 1.4 GZ flux density $S_{1.4} > 10$ mJy.  
The solid line is the OLS Y on X regression
line of equation 1.  The dotted line shows 
the equation for $\sigma_*$ = FWHM [O III]/2.35.
}
\label{fig:o3}
\end{figure}

\subsection{Radio-Loud Sample and Measurements}

We can use the radio core dominance parameter, determined from radio maps  
of our sources, to estimate jet orientation in quasars.
The Shen et al. (2008) sample only includes $\sim$50 objects with 
radio detections in FIRST (Becker et al. 1995), many of which are 
not formally radio-loud.  Using the [O III] $\lambda$5007 FWHM
to predict $\sigma_*$, we constructed a much larger sample 
from the SDSS more suitable for our purposes.

We started with the Shen et al. (2011) SDSS DR7 quasar catalog
and selected objects that have $z < 0.75$ to ensure
good measurements of [O III] $\lambda$5007 and the $\lambda$5100 \AA\
continuum used for SE black hole mass scaling relationships.   
We also only selected objects with $i < 19.1$,
where SDSS is nearly complete, and for which the signal-to-noise
ratio is not too low to make good spectral measurements. 

Next, we cross correlated this sample with the NRAO VLA Sky Survey (NVSS,
Condon et al. 1998), keeping matches with 
flux density $S_{1.4} > 10$ mJy and within 30$\arcsec$ of the SDSS
position (as recommended by Kimball \& Ivezi{\'c} 2008).  
The selection criteria ensures radio measurements 
well above the detection limit of both NVSS and FIRST and 
reliable matches.  At lower redshifts, however, this flux density limit
does not fully exclude radio-quiet quasars.  An additional 
criterion of the Shen et al. (2011) radio loudness parameter
RL $> 10$ removes the handful of remaining radio-quiet objects.
We also required that the slope parameter of the continuum of
H$\beta$ region have $\alpha_{\lambda} < -0.5$ to eliminate any quasars
with significant optical reddening, which would make them 
appear more intrinsically radio-loud then they actually are.
Then we considered the position of the FIRST radio core compared 
to the optical SDSS position, removing objects from our sample
that had mismatches greater than 2$\arcsec$; when the FIRST radio
cores and the optical positions do not match, the radio core
flux density likely represents a peak, or significant contamination,
from extended radio emission, which would result in a misleading
measurement of radio core dominance.  
Finally, we visually inspected the FIRST maps of all the remaining
sources in order to identify extended radio lobes (see, e.g.,
Kimball \& Ivezi{\'c} 2008).

This selection procedure produced a final sample of 386 radio-loud
quasars.  We then measured the radio core dominance by computing the 
ratio of FIRST peak flux density to that of the extended emission
(adding all the extended radio components together).  We k-corrected
these values to 5 GHz rest frame under the assumption that 
radio cores have a flat spectrum ($S_{\nu} \propto \nu^0$) and that
extended radio emission has a steep spectrum ($S_{\nu} \propto \nu^{-0.7}$),
resulting in our final core dominance measurements expressed as log R.
While this procedure will yield reliable core dominance measurements
for the majority of our sample, we must provide some caveats.
As correctly pointed out and characterized by Jackson \& Browne (2013),
the spatial resolution of the FIRST survey (5$\arcsec$) is too low
to resolve some radio sources, particularly those at higher redshifts.
While our selection criterion of matching SDSS and FIRST positions helps,
there will still be a number of poorly resolved sources that may cause
problems, particularly in the case of Compact Steep Spectrum (CSS) sources
(e.g., O'Dea 1998).  Kimball \& Ivezi{\'c} (2008) estimate about 25\% contamination
from CSS and Gigaherz Peaked Sources (GPS) among compact radio sources.
We note that when using log R for orientation, this contamination will
create extra scatter among core-dominant sources but should not 
artificially create correlations that don't exist.  The systematic
uncertainties involved with equating log R with orientation angle
may dominate over formal measurement errors, which are typically
about 0.15 dex.

\begin{table*}
 \centering
 \begin{minipage}{140mm}
  \caption{Optical measurements for the radio-loud sample. The full table is available online.}
  \begin{tabular}{@{}ccccccccc@{}}
  \hline
SDSS Name & z & iMAG & logL5100 & FWHM$_{H\beta}$ & log M$_{H\beta}$ & log L/L$_{Edd}$ & FWHM$_{[O III]}$ & log\ $\sigma_{* [OIII]}$\\
& & (mag) & (erg s$^{-1}$) & (km $s^{-1}$) & (M$_{\odot}$) &  & (km s$^{-1}$) & (km s$^{-1}$) \\
\hline
000111.19$-002011.5$ & 0.5179 & $-24.10$  & $44.699\pm0.007$ & $5016.1\pm584.6$ & $8.66\pm0.1$ &  $-1.09$ & $470.8\pm35$ & $2.32\pm$0.08\\
005905.51+000651.6   & 0.7189 & $-26.00$  & $45.446\pm0.007$ & $4976.2\pm1037.7$ & $8.94\pm0.13$ & $-0.64$ & $925.9\pm62$ & $2.46\pm$0.08\\
005917.47$-091953.7$ & 0.6409 & $-24.16$ & $44.795\pm0.004$ & $3653.9\pm408.1$ & $8.43\pm0.1$ & $-0.77$ & $306.4\pm21$ & $2.24\pm$0.08\\
010644.15$-103410.5$ & 0.4677 & $-24.54$ & $44.907\pm0.006$ & $3355.4\pm142.6$ & $8.42\pm0.04$ & $-0.64$ & $500.1\pm30$ & $2.33\pm$0.07\\
012905.32$-005450.5$ & 0.7067 & $-24.95$ & $45.132\pm0.02$ & $3020\pm750$ &     $8.68\pm0.12$ & $-0.76$ & $355.6\pm39$ & $2.27\pm$0.10\\
013352.66+011345.1   & 0.3081 & $-23.39$ & $44.461\pm0.007$ & $4205.1\pm422.7$ & $8.39\pm0.09$ & $-1.06$ & $310.4\pm27$ & $2.22\pm$0.08\\
021125.07$-081440.3$ & 0.5371 & $-23.88$ & $44.645\pm0.009$ & $4672.5\pm640.7$ & $8.57\pm0.12$ & $-1.06$ & $237.1\pm32$ & $2.18\pm$0.11\\
021225.56+010056.1   & 0.5128 & $-24.43$ & $44.805\pm0.007$ & $4945.3\pm678.4$ & $8.7\pm0.12$ & $-1.03$ & $303.3\pm17$ & $2.23\pm$0.07\\
030210.95$-075209.4$ & 0.7338 & $-24.93$ & $44.966\pm0.104$ & $5380.6\pm1249.2$ & $8.56\pm0.12$ & $-0.8$ & $187.6\pm26$ & $2.14\pm$0.12\\
073320.83+390505.1   & 0.6637 & $-25.14$ & $45.088\pm0.019$ & $2758.7\pm265.5$ & $8.34\pm0.08$ & $-0.38$ & $697.7\pm70$ & $2.40\pm$0.10\\
\hline
\end{tabular}
\end{minipage}
\end{table*}

\begin{table*}
 \centering
 \begin{minipage}{140mm}
  \caption{FIRST 1.4 GHz measurements for the radio-loud sample and core dominance. The full table is available online.}
  \begin{tabular}{@{}ccccc@{}}
  \hline
SDSS Name & z & Core & Extended & K-corrected log R\\
& & (mJy) & (mJy) & \\
\hline
000111.19$-002011.5$ & 0.5179 & 29.0 & 29.0 & 0.30\\
005905.51+000651.6 & 0.7189 & 2434.0 & 2434.0 & 1.19\\
005917.47$-091953.7$ & 0.6409 & 33.5 & 33.5 & 1.02\\
010644.15$-103410.5$ & 0.4677 & 269.0 & 269.0 & 0.28\\
012905.32$-005450.5$ & 0.7067 & 11.4 & 11.4 & 0.66\\
013352.66+011345.1 & 0.3081 & 16.7 & 16.7 & $-0.35$\\
021125.07$-081440.3$ & 0.5371 & 32.9 & 32.8 & 1.18\\
021225.56+010056.1 & 0.5128 & 51.1 & 51.1 & $-0.02$\\
030210.95$-075209.4$ & 0.7338 & 455.5 & 455.3 & 0.99\\
073320.83+390505.1 & 0.6637 &  130.5 & 130.5 & 0.93\\
\hline
\end{tabular}
\end{minipage}
\end{table*}

\begin{table*}
 \centering
 \begin{minipage}{140mm}
  \caption{Statistical description of the radio-loud sample}
  \begin{tabular}{@{}rrlc@{}}
  \hline
Property & Mean & SEMean & Standard Deviation  \\
\hline
z	&  0.51  & 0.008  &  0.15  \\
iMag   &  $-$24.45  &  0.046 &  0.91  \\
log L$_{5100}$  &  44.85  &  0.02 &  0.37  \\
log FWHM H$\beta$   &  3.73  &  0.01 & 0.23  \\
log M$_{H\beta}$  &  8.80  &  0.03   &  0.52  \\
log M$_{\sigma_*}$ & 7.92 & 0.02 & 0.38 \\
log L/L$_{Edd}$  & $-$1.08  & 0.02   &  0.46  \\
log FWHM [O III]  & 2.59  &  0.007  &  0.15  \\
log $\sigma_{* [OIII]}$ & 2.25  &  0.003  &  0.07 \\
log R  &  0.55  &  0.035  & 0.69  \\
\hline
\end{tabular}
\end{minipage}
\end{table*}

Table 2 gives the optical properties of our radio-loud sample,
many provided by the catalog Shen et al. (2011). 
The measurements of the FWHM [O III] profile are new and 
were done in the same manner as in \S 2.1.  The estimate of 
$\sigma_*$ uses equation 1.
Table 3 provides core and extended radio properties from the 
FIRST Survey as described above, as well as our adopted radio
core dominance parameter k-corrected to 5 GHz rest-frame. 
Table 4 provides some statistical properties of the tabulated parameters
that may be of interest.
  
\section{Analysis and results}

Our primary interest is using the radio core dominance log R in order 
to test how quantities involved in the M-$\sigma_*$ relation, as 
computed using quasar proxies, depend on orientation.  Some dependence,
especially on black hole mass computed from H$\beta$, is expected, and
likely contributes significant scatter.  It may also be possible to 
use ratios of black hole mass computed with orientation-dependent and 
orientation-independent quantities to estimate orientation in quasars,
whether or not radio loud.
We also note that we are using a moderately large sample of radio-loud quasars that may turn
up previously unrecognized rare objects.

Table 4 presents the Spearman rank correlation matrix for quantities of interest;
Pearson correlation coefficients give similar results.
To the quantities in tables 2 and 3, we add M$_{\sigma_*}$, the black hole 
mass computed from our $\sigma_{* [OIII]}$ using the M-$\sigma_*$ relationship
from McConnell \& Ma (2013), log M$_{\sigma}$ = 8.32 + 5.64 log($\sigma$/200 km s$^{-1}$), 
which we use to create a normalized mass M$_{H\beta}$/M$_{\sigma_*}$.
Below each correlation coefficient we give the two-sided probability of such 
a value arising by chance.  Many highly significant correlations are present,
and some notable non-correlations.
  
\begin{table*}
 \centering
 \begin{minipage}{140mm}
  \caption{Spearman Rank Correlation Coefficients and Probabilities}
  \begin{tabular}{@{}lcccccccc@{}}
  \hline
        & z & L$_{5100}$ & FWHM H$\beta$ & M$_{H\beta}$ &  L/L$_{Edd}$ & FWHM [O III] or $\sigma_*$ & M$_{H\beta}$/M$_{\sigma_*}$ \\
\hline
 L$_{5100}$ & 0.590 & & & & & & \\
         & $<$0.0001 & & & & & & \\
FWHM H$\beta$ & 0.093 & 0.155 & & & & & \\
          & 0.068 & 0.002 & & & & & \\
M$_{H\beta}$ & 0.303 & 0.496 & 0.898 & & & & \\
   &    $<$0.0001 & $<$0.0001 & $<$0.0001 & & & & \\
L/L$_{Edd}$  & 0.090 & 0.173 & $-$0.918 & $-$0.716 & & & \\
            & 0.0786 & 0.0006 & $<$0.0001 & $<$0.0001 & & & \\
FWHM [O III] or $\sigma_*$ & 0.074 & 0.192 & 0.167 & 0.225 & $-$0.079 & & \\
           & 0.147 & 0.0001 & 0.001 & $<$0.0001 & 0.12 & & \\
M$_{H\beta}$/M$_{\sigma_*}$ & 0.222 & 0.319 & 0.724 & 0.773 & $-$0.619 & $-$0.382 & \\
           & $<$0.0001 & $<$0.0001 & $<$0.0001 & $<$0.0001 & $<$0.0001 & $<$0.0001 & \\
log R & $-$0.088 & $-$0.061 & $-$0.271 & $-$0.248 & 0.248 & 0.003 & $-$0.203 \\
       & 0.086 & 0.235 & $<$0.0001 & $<$0.0001 & $<$0.0001 & 0.959 & $<$0.0001 \\
\hline
\end{tabular}
\end{minipage}
\end{table*}

An inspection of the correlation matrix shows that M$_{H\beta}$ and $\sigma_*$,
determined by their proxies, are significantly correlated, but the M-$\sigma$ relationship
in this sample needs a closer look.  
Figure 2 plots log M$_{H\beta}$ against log $\sigma_*$ for the radio-loud sample
(filled circles), and, in order to cover a wider range in parameter space to put our work
in context, the objects from Shen et al. (2008) we used to develop 
equation 1 (open circles) using self-consistent measurements.  
We also show a line fit for the local quiescent galaxy M-$\sigma_*$ relation (McConnell \& Ma 2013).
Our selection criteria lead to the sample containing most luminous and massive systems at z $<$ 0.75,
which are not representative of the full range of quasar properties.
The points for our sample sit well above the local M-$\sigma$ relationship, consistent
with the work of Gaskell (2009b) and Salviander \& Shields (2013) who used a similar approach employing 
similar proxies for mass and stellar velocity dispersion.  They also found
the most massive black holes shifted above the quiescent galaxy line.
Being rare and luminous objects, our radio-loud quasars also sit at higher redshifts
than the Shen et al. (2008) sample.
The M-$\sigma$ relation for quasars in our individual and combined samples changes with $z$, 
as Salviander \& Shields (2013) also found, and likely for the same reasons, 
involving a variety of selection biases, although they do not absolutely rule out true evolution.

\begin{figure}
\begin{center}
\includegraphics[width=8.9 truecm]{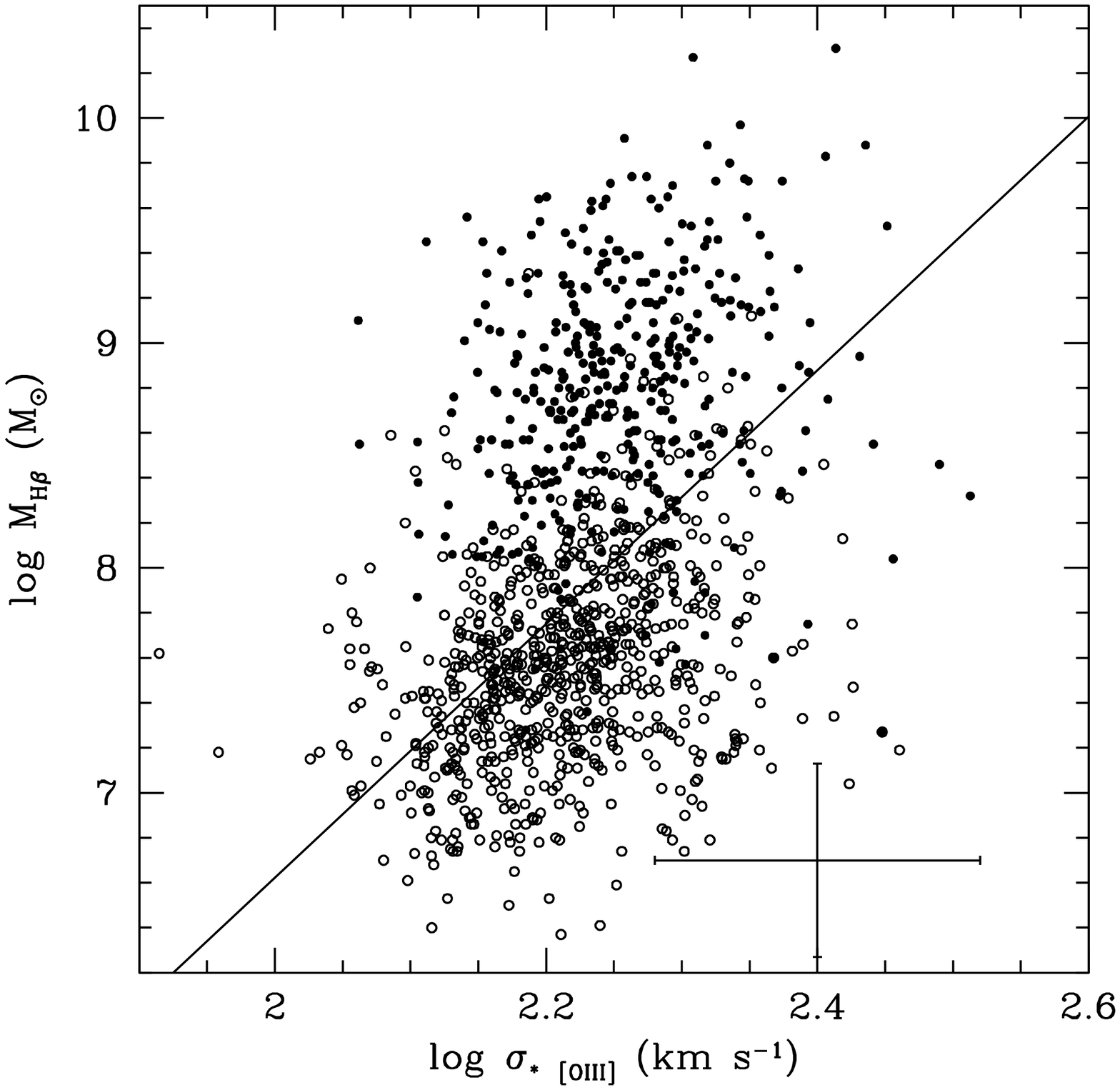}
\end{center}
\caption{
We plot log M$_{H\beta}$ against log $\sigma_*$ relation for our radio-loud sample (filled circles)
using estimates of these parameters from the SE black hole mass
scaling relationship using H$\beta$ (VP06) and equation 1.
The representative error bar displayed represents the
scatter in the predictive equations rather than measurement
errors, which are very much smaller.
For reference, we have also plotted the objects from Figure 1 (open
circles),
which are on average lower redshift and luminosity, using for consistency 
the same proxies to estimate mass and velocity dispersion.
The line shown is a recent update for the local M-$\sigma_*$ relation in
quiescent galaxies (McConnell \& Ma 2013).
}
\end{figure}

Next, Figure 3 reproduces the log R -- FWHM H$\beta$ plot of Wills \& Browne (1986),
again finding these quantities highly correlated in the sense that the more
edge-on sources have the broadest lines on average.  We note several differences,
however.  In particular, the upper right part of the figure is empty in 
Wills \& Browne (1986) but has a small population of objects in our sample.  
This likely arises in part from contamination from CSS sources, but 
some of these quasars, when checked in NASA's Extragalactic Database (NED), 
have radio spectra that are flat, not steep.  
We also note that our sample has few sources with log R $< -1$,
likely a result of excluding sources without close SDSS-FIRST core matches as well as 
the relatively high-frequency selection of FIRST and NVSS at 1.4 GHz; 
Wills \& Browne (1986) use quasars mainly selected from low-frequency surveys that
yield more lobe-dominant objects. 

\begin{figure}
\begin{center}
\includegraphics[width=8.9 truecm]{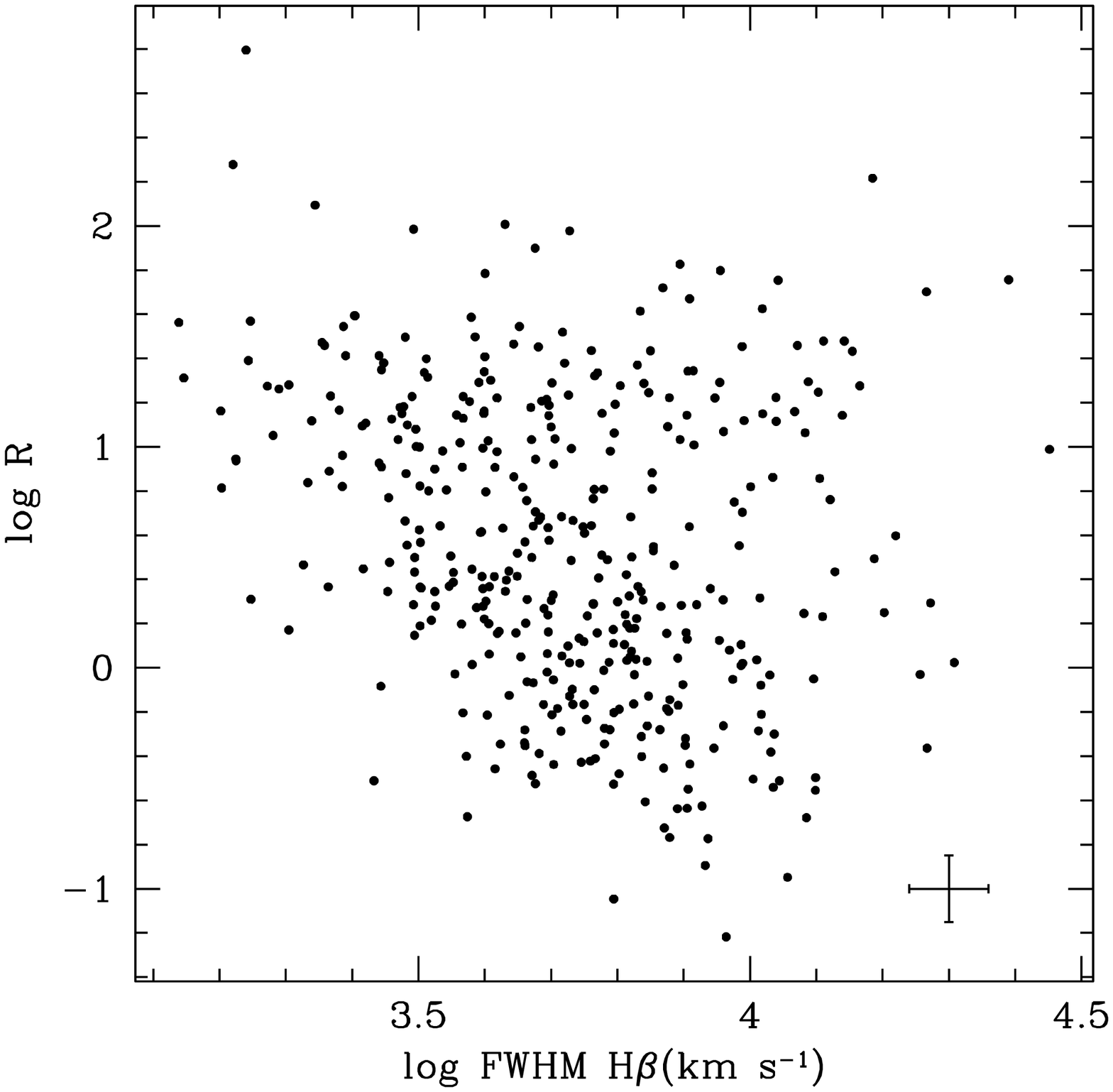}
\end{center}
\caption{
Log R is plotted against log FWHM H$\beta$.  The uncertainties
on log R are likely dominated by systematic issues as discussed
in the text and are not always well determined, while the uncertainties
on log FWHM H$\beta$ are $\sim$0.04 dex, very small compared
to the sample parameter range.
}
\end{figure}

Figure 4 shows a plot of log of the mass ratio M$_{H\beta}$/M$_{\sigma_*}$ 
against log FWHM H$\beta$.  This may be compared to a similar figure
by Shen \& Ho's (2014), also Figure 4, where they plot the ratio of
M$_{\sigma_*}$ to the H$\beta$ virial product, the so-called ``f-factor''
used to empirically correct H$\beta$-based masses for unknown
geometric effects.  
The scatter in M$_{H\beta}$/M$_{\sigma_*}$ for a given FWHM is roughly
consistent with that expected from the scatter in the proxies used
to construct the ratio.
Unlike Shen \& Ho (2014), we do not see
statistically significant improvement in the correlation if we 
create subsamples by binning in M$_{\sigma_*}$, perhaps because of 
our different parameter space and/or additional scatter from
using FWHM [O III] as a proxy.
In principle, M$_{\sigma_*}$ should normalize
the H$\beta$-based mass, as neither FWHM [O III]
nor the predicted $\sigma_*$ are significantly
correlated with log R, consistent with Brotherton (1996). 
We note that the average of the mass ratio is
not unity, as we have already shown how the most massive quasars
sit systematically above the local/quiescent galaxy M-$\sigma_*$ relationship
used to determine M$_{\sigma_*}$.
If the M-$\sigma-*$ relationship is consistent for this sample,
then mass variations cannot drive the correlation.  The 
uncertainties on FWHM H$\beta$, which enter into quantities on both axes, are too  
small to drive the correlation.
We therefore conclude, like Shen \& Ho (2014) whose sample was primarily 
radio-quiet quasars, that a major part of the variation in FWHM H$\beta$ arises 
from orientation effects, and that it is reflected in the M$_{H\beta}$/M$_{\sigma_*}$
ratio.

Table 5 shows a highly significant correlation ($\sim 5 \sigma$)
between M$_{H\beta}$/M$_{\sigma_*}$ and log R, which we plot
in Figure 5.  There is significant scatter, and caveats about
the possible intrinsic variation in the M-$\sigma_*$ relation
apply, but we can again use the OLS Y on X regression line to
predict log R from the orientation-biased mass ratio:

\begin{equation}
{\rm log\ R} = 0.797(\pm0.0615) - 0.278(\pm0.0584) {\rm log\  M}_{H\beta}/{\rm M}_{\sigma_*}
\end{equation}

Because of the large scatter, we would caution against using this
relationship for individual quasars.  Furthermore, because 
of the contamination of some fraction of expected CSS sources in 
the upper right portion of the distribution, which would
have much smaller log R values if measured with a survey
with higher spatial resolution than FIRST, the slope of the above
equation is likely steeper in reality.  VLA A-array observations
of this sample would be helpful to correct for 
this contamination in a future investigation. 

\begin{figure}
\begin{center}
\includegraphics[width=8.9 truecm]{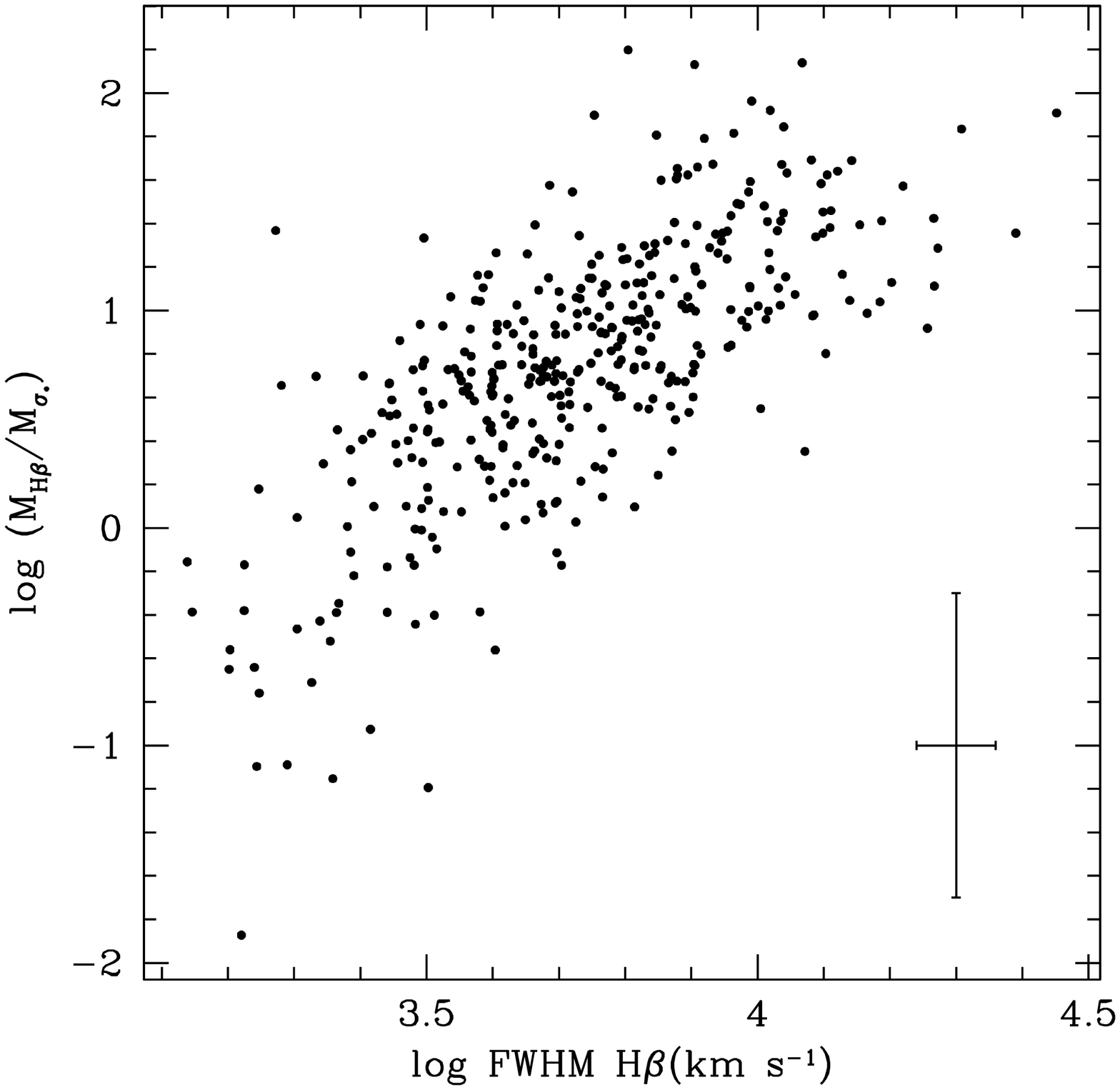}
\end{center}
\caption{
The log of M$_{H\beta}$/M$_{\sigma_*}$ is plotted 
against log FWHM H$\beta$.  The typical error bar 
is shown, where the magnitude for the y-axis 
represents scatter in the predictive relationships,
and in the x-axis measurement errors.
}
\label{fig:o3}
\end{figure}

\begin{figure}
\begin{center}
\includegraphics[width=8.9 truecm]{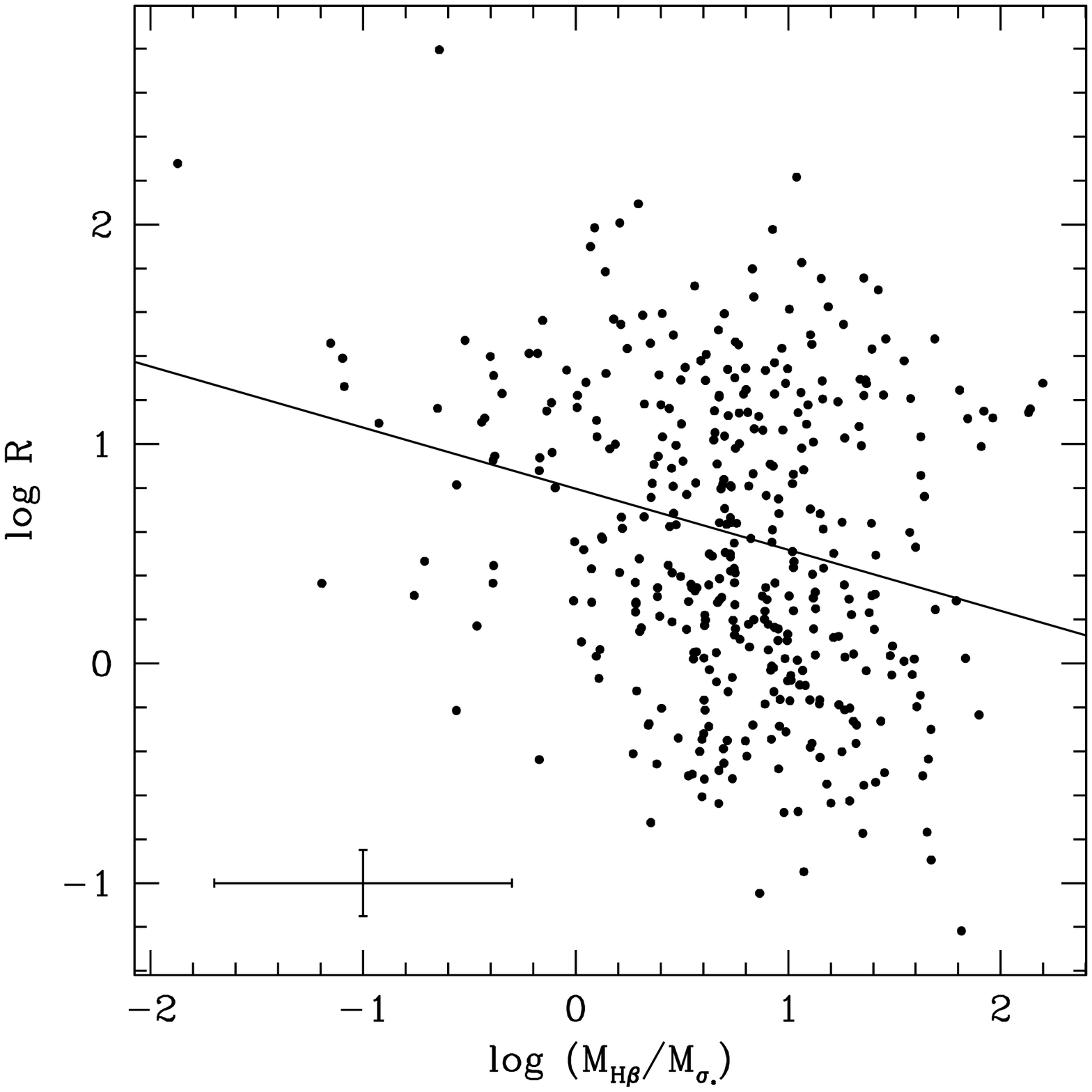}
\end{center}
\caption{
The log of M$_{H\beta}$/M$_{\sigma_*}$ is plotted
against log R, with uncertainties on M$_{H\beta}$/M$_{\sigma_*}$ 
again coming from scatter of the predictive equations.
The Spearman rank correlation coefficient is $-0.204$ with
a two-tailed probability of arising by chance of $<0.01\%$.
The solid line is the OLS Y on X regression line that
may be used to predict log R from the orientation-biased
to unbiased mass ratio.
}
\label{fig:o3}
\end{figure}

Finally, we performed one more experiment to test
the orientation dependence of M$_{H\beta}$/M$_{\sigma_*}$.
We computed a corrected black hole mass, using
equation 5 from Runnoe et al. (2013), that removed
the orientation dependence from M$_{H\beta}$ by
adding 0.173 $\times$ log R. Figure 6 repeats 
Figure 5, but uses this orientation-corrected black hole mass.
Although there is still similar scatter, the accuracy is
improved by removing orientation effects as
the Spearman rank correlation coefficient is reduced from
a highly significant $-$0.203 (two-tailed probability $< 0.01\%$ 
of arising by chance) to an insignificant $-0.0004$ 
(two-tailed probability of 99\% of arising by chance).  
This test uses a different sample than
Runnoe et al. (2013), and also a different way of normalizing 
the black hole mass (based on $\sigma_*$ rather than 
the C IV $\lambda$1549 line), represents independent
support for their mass correction term using log R.
While it has been clear for some time that H$\beta$-based
black hole mass determinations possess an orientation
bias, we have now quantified the effect and provided 
additional evidence that we can correct for it, at least statistically.

\begin{figure}
\begin{center}
\includegraphics[width=8.9 truecm]{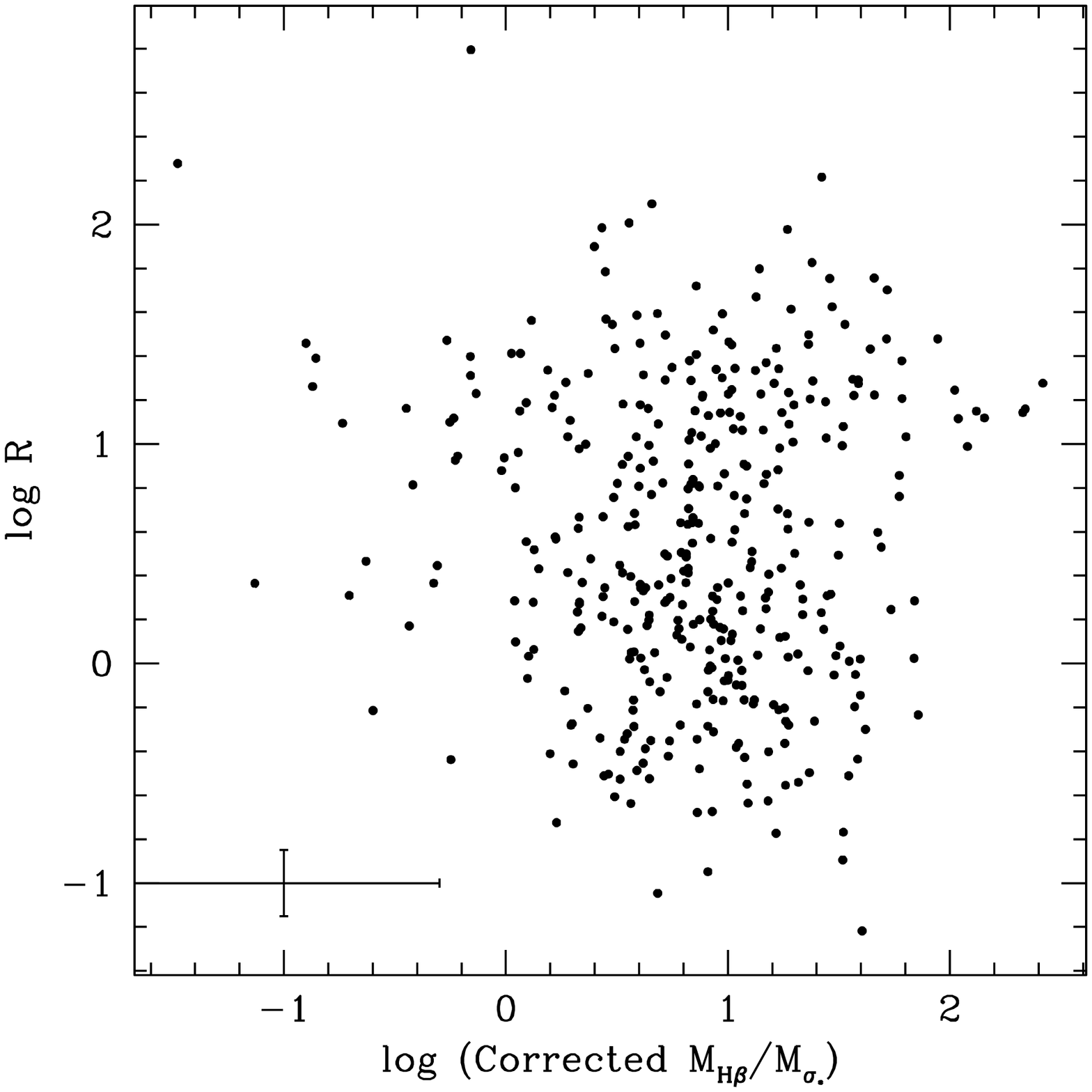}
\end{center}
\caption{
The log of M$_{H\beta}$/M$_{\sigma_*}$ is plotted
against log R, although this time M$_{H\beta}$
has been corrected for orientation bias 
using equation 5 from Runnoe et al. (2013).
The Spearman rank correlation coefficient is $-$0.0004,
with a two-tailed probability of arising by chance of 99\%,
indicating the variables are not correlated.
}
\label{fig:o3}
\end{figure}

\section{Discussion}

Previous work and our results establish
that black hole masses determined
using the broad ${H\beta}$ emission line suffer 
orientation bias, certainly in radio-loud quasars
and very likely in radio-quiet quasars as well
(e.g., Krolik 2001; Collin et al. 2006).  Based
on the investigation of Runnoe et al. (2013),
this is also true using alternative velocity measurements
like the line dispersion 
$\sigma_{line}$ and true for Mg II $\lambda$2800, which
behaves similar to H$\beta$.  Quantitatively the bias is
close to an order of magnitude over the observed 
full range of quasar orientations, and represents a significant
amount of scatter in many applications
(e.g., studying M-$\sigma_*$ as in Salviander \& Shields 2013).
There is no reason to think that reverberation
mapping black hole masses are immune from orientation
effects, although this would be good to confirm observationally.

Orientation also affects optical luminosity (e.g., Nemmen \& Brotherton 2010,
Runnoe et al. 2013, and references therein), which can 
manifest in numerous ways.  For instance, 
flux-limited samples possess strong orientation
biases that can change the slope and normalization 
of quasar luminosity functions (e.g., DiPompeo et al. 2014). 
We agree with the conclusions of Shen \& Ho (2014)
that orientation is an important parameter
affecting the appearance of type 1 quasars,
above and beyond being a primary reason for type 1 and type 2
AGNs more generally (e.g., Antonucci 1993).

As a practical matter, having multiple estimators
for quasar black hole mass, that can be computed 
without radio observations and that have different
dependencies on orientation, offers the promise of 
an orientation indicator that can be used in all
quasars, radio-loud and radio-quiet alike.
Such a general orientation indicator could be 
used to look for orientation bias, as well as to correct for its
effects.

The ratio of an orientation-biased black hole mass
to one derived from an unbiased stellar velocity dispersion $\sigma_*$, or
a reasonable proxy like FWHM [O III], can serve as a rough
orientation indicator.  There remain issues to consider besides
the scatter.  The intrinsic
M-$\sigma_*$ ratio may vary with AGN type, the host galaxy type,
the black hole mass, and perhaps redshift as well (in addition
to discussions by Gaskell 2009b and Salviander \& Shields 2013, see also 
Bennert et al. 20011, Kormendy \& Ho 2013, Woo et al. 2015, and others).
One additional issue we propose here is an orientation effect
with luminosity.  If ideas about a receding torus and 
a correlation between luminosity and opening angle are correct
(e.g., Lawrence 1991; Ma \& Wang 2013), then more luminous quasars
will be on average seen at larger viewing angles.  This in
turn will mean that they will on average have larger FWHM H$\beta$
and therefore correspondingly larger black hole mass estimates
when using H$\beta$ single-epoch scaling relationships.
This effect could contribute to more luminous quasars
also sitting higher in M-$\sigma$ diagrams like our Figure 2.
These topics are of course of great interest
and study in their own right.  

Because of the large amount of scatter between these parameters,
this predictive equation is unlikely to be reliable for determining 
orientation in individual objects.  However, given the large numbers 
of quasars now being routinely studied, it may prove useful for
statistical studies.  There may also be opportunities to improve
the relationship and reduce scatter.

We previously discussed some caveats associated with
the use of the FIRST survey to determine radio core 
dominance.  Certainly there are some compact steep
spectrum quasars that would have small values of log R
if observed at higher spatial resolution, but 
are interpreted as jet-on quasars with large core
dominance.  This effect creates extra scatter in Figures 
3 and 5.  A fraction of our sample has
additional radio measurements available from the Westerbork
Northern Sky Survey at 325 MHz (WENSS, Rengelink et al. 1997).
Some of the objects with both large log R and FWHM H$\beta$
that sit in the upper right corner of Figure 3, an
empty area in the plot of Wills \& Browne (1986),
do indeed have steep radio spectra.  
Some, however, have flat spectra and probably do indeed represent jet-on systems.

If the scenario of Wills \& Browne (1986), proposing that
H$\beta$ is emitted from a disc-shaped BLR in the plane
perpendicular to the jet axis, is right, how could 
some of these jet-on systems show the broadest H$\beta$
lines of all?  For example, SDSS J161826.93+081950.7
has log R = 1.76 and a flat radio spectrum, but 
also FWHM H$\beta$ = 24,000 km s$^{-1}$, one of
the broadest broad emission lines ever seen in a 
quasar.  Suggestive is the fact that the [O III]
profile has a bump on the blue side, indicating
that it is perhaps double-peaked, which has been
previously noted by, e.g., Smith et al. (2010) and
Fu et al. (2011), who included the quasar as a 
binary candidate.  We propose that these rare 
objects, with apparently misaligned jet and BLR
axes, could result from recent mergers. 
That is, a merger has already occurred and 
changed the system spin axis seen on large 
scales vs. small scales.  Thus we see the 
large scale radio structure as indicating
a jet-on orientation, but the broad line region
is more edge-on. 
Follow-up investigations of the jets on
millarcsecond scales with VLBI would be of interest.

\section{Summary}

We recalibrated FWHM [O III] as a proxy 
for $\sigma_*$ using the sample and measurements 
of Shen et al. (2008).  We then constructed
a sample of radio-loud quasars based on a 
combination of SDSS, NVSS, and FIRST catalogs
and measured FWHM [O III] in order to predict
$\sigma_*$.  We found that single-epoch 
scaling relationship estimates of black hole mass
using H$\beta$ appear to be strongly correlated 
with radio-based orientation indicators, even
when normalized by mass determinations from 
the M-$\sigma_*$ relation, consistent 
with some previous work.  We discussed 
using the ratio of biased quasar black hole mass
to stellar velocity dispersion as an orientation
indicator, as well as a small population of 
outliers with both large log R and FWHM H$\beta$,
which may represent recently merged systems.

\section*{Acknowledgments}

We thank Zhaohui Shang and Bev Wills for useful discussions, and 
the referee Martin Gaskell for helpful suggestions.
Support for Vikram Singh came from Wyoming EPSCoR Undergraduate Research Fellowship Program and NSF Grant EPS-1208909.

\label{lastpage}

\end{document}